\documentclass[conference]{IEEEtran}

\usepackage[
backend=biber,
style=numeric,
sorting=none
]{biblatex}
\usepackage{amsmath}
\usepackage{algorithmic}
\usepackage{booktabs}
\usepackage{comment}

\usepackage{pgfplots}
\usepackage{pgfplotstable}
\pgfplotsset{compat=1.17}

\usepackage{multirow}
\usepackage{array}

\usepackage{xcolor}
\definecolor{redbullet}{RGB}{234,67,53}
\definecolor{bluebullet}{RGB}{66,133,244}
\definecolor{yellowbullet}{RGB}{251,188,4}
\definecolor{greenbullet}{RGB}{52,168,83}

\def\BibTeX{{\rm B\kern-.05em{\sc i\kern-.025em b}\kern-.08em
    T\kern-.1667em\lower.7ex\hbox{E}\kern-.125emX}}

\begin{document}

\title{Exploring Correlation Patterns in the Ethereum Validator Network}

\author{
  \IEEEauthorblockN{Simon Brown}
  \IEEEauthorblockA{\textit{ConsenSys Software Inc.}\\
  simon.brown@consensys.net}
  \and
  \IEEEauthorblockN{Leonardo Bautista-Gomez}
  \IEEEauthorblockA{\textit{MigaLabs}\\
  leobago@protonmail.com}
}

\maketitle

\begin{abstract}
There have been several studies into measuring the level of decentralization in Ethereum through applying various indices to indicate the relative dominance of entities in different domains in the ecosystem.  However, these indices do not capture any correlation between those different entities, that could potentially make them the subject of external coercion, or covert collusion.  We propose an  index that measures the relative dominance of entities based on the application of correlation factors.  We posit that this approach produces a more nuanced and accurate index of decentralization.
\end{abstract}


\section{Introduction}

There have been several attempts to model a heuristic to measure the level of decentralization in the Ethereum ecosystem, that have relied on various techniques and indices that have been borrowed from the fields of economics and ecology \cite{wu2020coefficient} \cite{gupta2018gini} \cite{gochhayat2020measuring} \cite{lee2021dq}.  These include indices such as the Gini index and the Herfindahl–Hirschman Index (HHI), as well as several indices that are derived from the measurement of entropy \cite{brown2023measuring}.  When used in combination, these measurements reveal useful insights into the relative market share and/or share of resources of various entities, which can indicate the areas of concentration or diffusion of control and influence in the ecosystem.

While these measurements can prove useful in measuring decentralization at a high level, they fail to capture the nuance in the correlation between various entities within the ecosystem, which can potentially lead to subtle implicit collusion and/or potential coercion by external actors.  We therefore propose a model that seeks to capture the level of correlation between entities in the ecosystem across a number of dimensions, and present our findings from applying the model to available network data.  We demonstrate how the level of correlation between independent entities can reduce the effective level of decentralization in certain cases, while increasing the effective level of decentralization in other cases.

This paper is organized as follows: in section~\ref{sec:background}, we discuss the background for this research; in section~\ref{sec:motivation}, we describe the motivation for this work. The data used for this work is presented in section~\ref{sec:data}. Our methodology and the various calculations used in our model are presented in section~\ref{sec:methodology}.  In section~\ref{sec:results}, we outline the results of applying our model to the underlying datasets. We summarize the results and learnings in section~\ref{sec:discussion}, and finally discuss future work and areas for further research in section~\ref{sec:conclusion}.

\section{Background}
\label{sec:background}

Ethereum currently has over 916,000 validators \cite{beaconchain2024} attached to approximately 5,000 to 6,000 nodes, out of a total of approximately 12,000 nodes across the entire network \cite{nodewatch2024}.

Many validator clients are controlled by staking pools, with only about 25\% of the validator set being independent solo stakers \cite{dune2024}.  Several staking pools have garnered a relatively large market share, and some employ a number of different node operators to run nodes on the network, to which validators are attached.  Node operators can attach any number of validators to the nodes they run, and can employ any combination of execution client and consensus client, of which there are approximately 6 widely adopted clients of each available.

Furthermore, node operators may choose to connect to independent, third-party block builders to source the blocks that the validators attached to their nodes propose to the network. Nodes connect to builders via relayers in the mev-boost system.  Relayers can be run by independent providers or by the same institutions offering MEV through block builders. Relayers provide a quasi-escrow service for block builders and validators to negotiate payment and delivery for blocks.

\section{Motivation}
\label{sec:motivation}

At a high level, we can observe patterns within the ecosystem that highlight potential areas of concern, including potential implications for the network's security in terms of client diversity \cite{clientdiversity2024}, or an over-reliance on certain infrastructure providers (e.g. cloud providers, relayers etc.), as well as the network's ability to withstand coercion from regulatory overreach in any number of jurisdictions \cite{wahrstatter2023}. These insights do not necessarily reveal any indication of why these patterns of concern occur in the first place however.

An example of why the correlation between entities is important to measure is when considering the market share of staking pools. Different staking pools have different policies regarding node operators, including geographical location and client diversity \cite{vanom2024}.  Through segmenting the validator set by staking pool, we can measure the level of decentralization within each staking pool, by measuring the diversity of node operators, clients and relays within each staking pool, adjusted for correlation to each other. The resulting measurement can then be used to adjust the market share of staking pools to more accurately reflect the effective level of decentralization.  Similarly, while the market concentration of node operators is ostensibly diffuse, through measuring the correlation between node operators across several dimensions, we can start to see that a more accurate level of concentration than simply looking at the market share alone.

\section{Data Sources}
\label{sec:data}

Our study analyses three discrete datasets from two independent sources:

\begin{itemize}
    \item Dataset A: sourced from rated.network \cite{ratednetwork2024}, which pertains to node operators and staking pools, sourced on the 15th January 2024, and contains data from the preceding 30 days.
    \item Dataset B: sourced from MigaLabs, and which contains data pertaining to individual beacon nodes on the network.  This dataset includes the number of attestation subnets each node advertises.
    \item Dataset C: sourced from MigaLabs, and which contains data pertaining to individual beacon nodes on the network, but with a smaller sample size, less than 500 records. This dataset contains the specific number of validators attached to each node.
\end{itemize}

Each dataset contains a discrete set of attributes, which we analyze for patterns of correlation.  The attributes that are analysed in each dataset include:

\begin{itemize}
    \item \textbf{Dataset A} (Node Operators):
    \begin{itemize}
        \item Market share of node operator
        \item Percentage breakdown staking pools served
        \item Percentage breakdown of client software
        \item Percentage breakdown of relayers used
    \end{itemize}
    \item \textbf{Dataset B} (Individual Nodes):
            \begin{itemize}
                \item Country of Operation
                \item Consensus Client Software
                \item ISP / Datacenter
                \item Number of advertised attestation subnets
            \end{itemize}
    \item \textbf{Dataset C} (Individual Nodes):
            \begin{itemize}
                \item Country of Operation
                \item Consensus Client Software
                \item ISP / Datacenter
                \item Number of validator clients
            \end{itemize}
\end{itemize}

In order to produce datasets B and C, we leverage a network crawler that was developed by MigaLabs, which constantly connects to beacon nodes in the p2p network and keeps track of their metadata, which they share as part of the initial handshake. Beacon nodes share information about the attestation networks they are interested in, their connection addresses (IPv4, IPv6, ports, etc.), and sometimes the consensus client they use. Using external APIs, we obtain country and ISP information for each beacon node, as this is easily obtainable once you have the IP address of the node.

Beacon nodes do not advertise which validators they run. Knowing which specific operator runs a node could open targeted attack vectors (e.g., DoS) that could affect the stability of the network. Instead, the attestation networks a node is interested in can be taken as an indicator of the number of validators running behind the node.

\begin{figure}[htbp]
    \centering
    \includegraphics[width=0.95\linewidth]{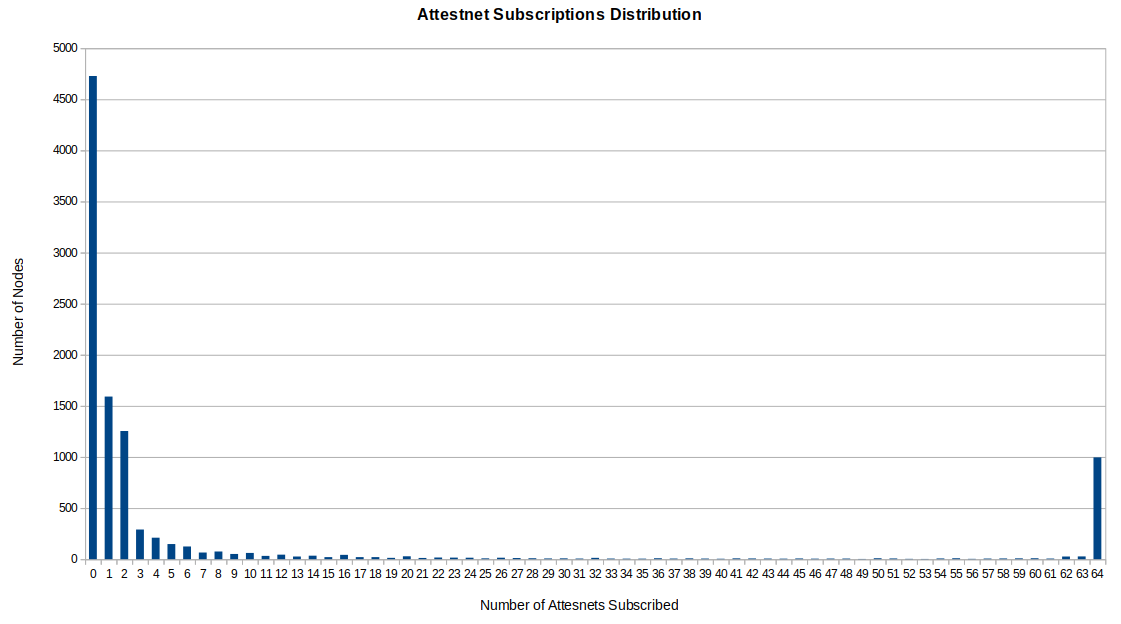}
    \caption{Attesnet Distribution Dataset B}
    \label{fig:attesnets}
\end{figure}

We see in Figure~\ref{fig:attesnets} the distribution of attesnet subscriptions at the time Dataset B snapshot was taken. We see almost 5,000 nodes with zero attesnets, which are likely to be nodes without validators behind. We also see thousands of nodes with a small number of attesnets, which are likely to be solo-stakers. Then, on the right side of the figure, we see nodes registered to 64 attesnets, the maximum nodes can register to. For nodes following all 64 attestation networks, it becomes hard to know whether they are running 64 or a thousand validators. 

There are several ways to know the number of validators running behind a set of nodes. The most obvious one is to ask. Large staking pools (e.g., Lido) might ask their node operators to report on this type of infrastructure choice. Also, there exist some techniques, such as network message triangulation, that could allow a set of nodes to have a good vision of how many validators are run behind a node. These techniques are costly to run, and it is extremely difficult to get a detailed vision of the entire network. Nonetheless, they can be used to extract partial knowledge of a small number of nodes in the network. There are also some network players (e.g., relays) that have direct communication with validators, allowing them to gather more information about the nodes in the network.

An explanation of message triangulation, and other techniques to extract data from the network, are out of the scope of this paper. However, these techniques were employed to create a dataset, which is a sample of nodes on the network in which the exact number of validators attached to each node is known. The validator distribution observed in this random sample of nodes is used to estimate the distribution of the number of validators attached to nodes that advertise 64 attestation subnets in Dataset B. We also cross-filter this random sample of nodes with the data from the Armiarma crawler to create Dataset C, which we use in calculations in section \ref{sec:herfindahl–hirschman-indices-for-individual-nodes}.

\section{Methodology}
\label{sec:methodology}

Our analysis applies several calculations to each dataset in order to attempt to identify any correlations between entities in the dataset across various attributes, and any correlations between specific attributes.

We describe a novel index calculation, based on the Herfindahl–Hirschman Index, but with an additional correlation factor, to account for potential correlations between staking pools and node operators in section~\ref{sec:modified-hhi}.

We examine the level of correlation between variability in attributes in dataset A in section~\ref{sec:correlation-between-variability-in-attributes}, in order to identify any correlation between a staking pool and the level of client diversity, or the range of node operators or relayers that are used by that staking pool.

We attempt to calculate a correlation between the market share of node operators and the consensus clients they run, and relayers that they use, by mapping each percentile of market share to the total number of clients / relays used by staking pools and node operators in that percentile in section~\ref{sec:correlation-between-operators-size-clients-and-relayers}.

Our analysis also examines individual nodes on the network by calculating any correlation between individual nodes in dataset B using standard statistical measurements of correlation in section~\ref{sec:calculating-correlation-between-individual-nodes}.  We also employ a novel approach for finding correlation between attributes in section~\ref{sec:ranking-the-level-of-correlation-between-attributes}.

\subsection{Standard Herfindahl–Hirschman Index}
\label{sec:standard-hhi}

Our model uses the Herfindahl–Hirschman Index as a basis for measuring the market concentration of staking pools and node operators.  The Herfindahl–Hirschman Index (HHI) is a widely adopted economic index for measuring market concentration across a number of economic sectors \cite{OECD2021}.

The HHI relies on market percentage shares, and is defined as the sum of the square of the percentage market share of each entity in a population.  It therefore results in a value close to 0 for a population in which each entity has a relatively equal share, but approaches $100^2$ in a population which is dominated by a relatively small number of entities.

\subsection{Modified Herfindahl–Hirschman Index}
\label{sec:modified-hhi}
This is calculated by summing the market share of every entity multiplied by the market share of every other entity and an additional correlation factor that indicates how correlated the respective entities are.  The correlation factor, $\rho_{ij}$, between entities $i$ and $j$, adjusts their respective market shares to account for their level of similarity across various attributes.

\[
HHI' = \sum_i \sum_j \left( s_i \cdot s_j \cdot \left( \rho_{ij} \cdot 100^{-n} \right) \right)
\]

Where $s_i$ and $s_j$ represent the market shares of entities $i$ and $j$ respectively, and $n$ is the number of attributes with which the correlation factor $\rho_{ij}$ is calculated. When applied to the set of staking pools derived from dataset A, wherein each entity is a unique staking pool, the correlation factor $\rho_{ij}$ is calculated for each pairwise comparison in the dataset as follows:

Let $R_i$, $C_i$ and $O_i$ be the sets of relays, clients, and operators for entity $i$, respectively. Similarly, let $R_j$, $C_j$ and $O_j$ be the sets of relays, clients, and operators for entity $j$.  The values in each set represent the percentage of each relay, client, or operator used by the respective entity, and therefore the sum total of all values in each set is 100.

The correlation factor $\rho_{ij}$ between entities $i$ and $j$ is defined as the sum of the minimum of each corresponding value in the sets $R_i$, $C_i$ and $O_i$ and the sets $R_j$, $C_j$ and $O_j$ respectively. This can then be expressed algebraically as follows:

\[
\sum_{k=1}^{|R|} \min(R_{ik}, R_{jk}) + \sum_{k=1}^{|C|} \min(C_{ik}, C_{jk}) + \sum_{k=1}^{|O|} \min(O_{ik}, O_{jk})
\]

\vspace{8pt}

In other words, the correlation factor is derived from examining the clients, relays and operators that each pair of staking pools have in common, and taking the minimum percentage that each staking pool uses in each case, and adding those percentages together.

We calculate and compare both the standard HHI and and the modified HHI for both staking pools and node operators.  In the case of node operators, the correlation factor only considers clients and relayers.

\subsection{Calculating correlation between variability in attributes}
\label{sec:correlation-between-variability-in-attributes}

Our analysis attempts to measure any correlation between the level of variability in clients, relayers and node operators across staking pools. This is done to try to identify if there is a correlation between the size of a staking pool and the level of client diversity, or the range of node operators or relayers that are used.

This is achieved by first calculating the coefficient of variance in the percentages of relayers, clients and node operators used by each staking pool. From this we obtain a matrix with a row for each staking pool, a column for market share, and columns for the coefficients of variation for clients, relayers and node operators.  We then calculate the $R^2$ value for each pairwise combination of columns to identify any potential correlation.

The coefficient of variation (CV) as a measure of relative variability is calculated as the ratio of the standard deviation ($\sigma$) to the mean ($\mu$) of a set of clients, relayers and node operators respectively, defined as:

\[ CV = \left( \frac{\sigma}{\mu} \right) \times 100 \]

Using the resulting matrix of staking pools and the coefficient of variation of their attributes: clients, relayers and node operators, we can then calculate the Pearson correlation coefficient \cite{pearson1895} for each pairwise combination of attributes.  The Pearson correlation coefficient, $r$, is well know to any student of statistics and is defined as:

\[
r = \frac{\sum{(X_i - \bar{X})(Y_i - \bar{Y})}}{\sqrt{\sum{(X_i - \bar{X})^2}\sum{(Y_i - \bar{Y})^2}}}
\]

\vspace{3pt}

Where $X_i$ and $Y_i$ are the individual data points of attributes $X$ and $Y$, and $\bar{X}$ and $\bar{Y}$ are the means of attributes $X$ and $Y$, respectively.  For each $r$ value, we square the result to obtain the coefficient of determination, $R^2$, which indicates any potential correlation between attributes.

We also apply the above analysis to node operators, whereby we attempt to analyze whether there is any correlation between variability in the market share of the node operators and the clients that they run or relayers they use.

\subsection{Correlation between operators size, clients and relayers}
\label{sec:correlation-between-operators-size-clients-and-relayers}

Our analysis also attempts to calculate a correlation between the market share of node operators and the consensus clients they run.

This  involves analyzing the set of node operators $K$, each possessing a market share represented by $m$, along with the percentage breakdown of the consensus clients they run, denoted as $c = { c_1, c_2 . . . c_n }$. Each $c_i$ signifies the percentage of a specific client in the known client set $C$, run by the operator $k_i$ in $K$.

A function is employed to construct a matrix, denoted as $M$, wherein each row corresponds to a market share percentile $d \in \left\{0, 1, ... 9\right\}$, and each column corresponds to a client $c \in C$. In this matrix, each column aggregates the sum of percentages of that client that is run by all node operators in the dataset possessing the respective market share percentile.

To generate this matrix, the function $f: m_i \mapsto d$, iterates through the set $K$. For each entity $k_i \in K$, it maps the entity's percentage market share $m_i$ to the corresponding percentile $d \in \left\{0, 1, ... 9\right\}$ and increases the values in the columns of $M$ corresponding to each client $c_j \in C$ by the percentage run by node operator $k_i$.

The resulting matrix provides a view of any correlation of market share of node operators to the clients they use, and the diversity of clients they use, visualized as:

\[
M = \begin{bmatrix}
m_1 & M_1^1 & M_1^2 & \ldots & M_1^{\nu} \\
m_2 & M_2^1 & M_2^2 & \ldots & M_2^{\nu} \\
\vdots & \vdots & \vdots & \ddots & \vdots \\
m_n & M_{\eta}^0 & M_{\eta}^1 & \ldots & M_{\eta}^{\nu}
\end{bmatrix}
\]

where $\nu$ is the size of the set of clients, and $\eta$ is the size of the set of node operators.

This analysis can indicate if any particular client is favoured by larger or smaller node operators. The process is repeated for relayers as well, whereby we try to establish any correlation between the market share of node operators and which relayers they use, as well as the number of relayers they use.

\subsection{Calculating correlation between individual nodes}
\label{sec:calculating-correlation-between-individual-nodes}

In order to determine if there is any correlation between individual nodes on the network across various attributes, we analysed dataset B, which contains data on individual nodes on the network, including country, client, ISP, and number of attestation subnets advertised.

\subsubsection{Calculating Chi-squared value between attributes}
\label{sec:calculating-chi-squared-value-between-attributes}

We analyzed the data by calculating the Chi-squared value between each pairwise attribute, we then calculated the corresponding p-value, and finally derived the Cramers-V value.

To do this we first generate a contingency table for each pairwise comparison of attributes in the dataset.  Each contingency table has rows for each value of attribute A, and columns for each value of attribute B, where each cell contains the frequency of occurrences for each combination of values of attribute A and B respectively. The Chi-squared value is then calculated as:

\[
\chi^2 = \sum \frac{(O_{ij} - E_{ij})^2}{E_{ij}}
\]

Where $O_{ij}$ is the observed frequency in cell $(i, j)$, and $E_{ij}$ is the expected frequency in cell $(i, j)$, calculated as:

\[
E_{ij} = \frac{(\text{row sum})(\text{column sum})}{\text{total number of nodes}}
\]

\subsubsection{Calculating Cramers-V value between attributes}
\label{sec:calculating-cramers-v-value-between-attributes}

We also calculate the Cramers-V value \cite{akoglu2018user} for each pairwise comparison as a complimentary measurement to the result of both the Chi-squared value and p-value of each pairwise comparison of attributes, in order to help to gauge the strength of any observed correlation.

Cramer's V is calculated using the following formula:

\[ V = \sqrt{\frac{\chi^2}{n \cdot \min(k-1, r-1)}} \]

Where:
\begin{itemize}
    \item $\chi^2$ is the chi-squared statistic obtained from the contingency table, as previously calculated,
    \item $n$ is the total number of observations in the table,
    \item $k$ is the number of columns in the table,
    \item $r$ is the number of rows in the table.
\end{itemize}

\vspace{8pt}

Cramer's V ranges from 0 to 1, where 0 indicates no correlation between the attributes, and 1 indicates they are totally correlated with one another.  While this allows us to determine if there is any correlation between the market share of node operators and the consensus clients they run, we also apply this analysis to size of node operators and relayers they use.

\subsection{Ranking the level of correlation between attributes}
\label{sec:ranking-the-level-of-correlation-between-attributes}

Our analysis also includes a function to determine which attributes display the highest amount of correlation between individual nodes in the network.

For every record in the dataset, we compare it to every other record in the dataset along a specific attribute, including country, client, ISP, and number of attestation subnets advertised.

Where the attributes are equal for each record, we record a 1, where they are not equal, we record 0.
The result is a bitstring for each attribute from which we can derive a hamming weight.
The process can be described as incrementing a count every time the attribute for each record being compared is equal, and is essentially a method for deriving a count for each unique value observed in a specific attribute. We then repeat the entire process for the next attribute.

The result is a series of hamming weights for each record compared to every other record, for each attribute. We then add all the hamming weights together for each record, so every record has an aggregate weight, resulting in a table with the record index in column 1, a column for the hamming weight of each attribute, and a column for the sum of all hamming weights for that record.

This aggregate weight indicates the level of correlation of the respective node to other nodes, and allows us to rank the dataset to help identifying patterns between correlations. The hamming weights in each column represent the sum of all observances of the value of the attribute of the respective record.

The process is defined formally as:

Let \( N \) be the number of records in the dataset, and \( M \) be the number of attributes.  For each pair of records \( i \) and \( j \) (\( i, j \in \{1, 2, \ldots, N\} \) and \( i \neq j \)), and for each attribute \( k \) (\( k \in \{1, 2, \ldots, M\} \)), we define a binary function \( \delta \left( i_k,j_k \right) \) using the Kronecker delta function:

\[
\delta \left( i_k,j_k \right) = \begin{cases} 1 & \text{if } i_k = i_k \\ 0 & \text{if } i_k \neq j_k \end{cases}
\]

The hamming weight \( H_{ik} \) for record \( i \) and attribute \( k \) is the sum of all binary values for that attribute across all pairwise comparisons:

\[ H_{ik} = \sum_{j=1, j\neq i}^{N} \delta \left( i_k,j_k \right) \]

The aggregate hamming weight \( A_{i} \) for record \( i \) is the sum of hamming weights across all attributes:

\[ A_{i} = \sum_{k=1}^{M} H_{ik} \]

The final table can be represented as a matrix \( T \) with \( N \) rows and \( M+2 \) columns, where the first column contains the record index, columns \( 2 \) to \( M+1 \) contain the hamming weights for each attribute, and the last column contains the aggregate hamming weight:

\[ T = \begin{bmatrix} 
1 & H_{1,1} & H_{1,2} & \ldots & H_{1,M} & A_{1} \\
2 & H_{2,1} & H_{2,2} & \ldots & H_{2,M} & A_{2} \\
\vdots & \vdots & \vdots & \vdots & \vdots & \vdots \\
N & H_{N,1} & H_{N,2} & \ldots & H_{N,M} & A_{N}
\end{bmatrix} \]

\vspace{2pt}

This matrix \( T \) represents a table with the record index, individual hamming weights for each attribute, and the aggregate hamming weight for each record.

\section{Results}
\label{sec:results}

We present the results from the application of each calculation described in the previous section to the dataset A and B.  The results are detailed in the relevant subsections that follow.  Discussion of the results and their possible interpretations, as well as any future work that the results suggest, is expanded upon in the conclusion section.

\subsection{Calculating correlation between variability in attributes}

Our analysis first examines the level of variability in the clients and relayers used by each node operator, as described in section~\ref{sec:correlation-between-variability-in-attributes}.  Each node operator is given a coefficient of variance for each attribute, calculated from the respective percentages of clients and relayers that the node operator uses. We then attempt to calculate any correlation between the variability in each attribute by calculating the $R^2$ value for each pair of attributes.  This method is also applied to the variability in clients, relayers and node operators that each staking pool uses.

\subsubsection{Correlation of variability across attributes for Node Operators}

The following table presents the $R^2$ values for comparison of variability between attributes in datasets A with respect to node operators. As can be seen from the results below, \textbf{there is strong correlation between the market share of the node operator and the variability in the distribution of relays they have procured blocks from}.

Overall the results from this specific analysis are inconclusive.  The high variability in relays may be partially explained by the fact solo stakers propose blocks relatively infrequently, and may have proposed only one or two blocks within the sample period, and this may be represented in the data in terms of which relayers they procured those blocks from.  In this context they may have connected to multiple relayers, but only proposed blocks that were sourced from one or two relayers within the sample period. In order to make the results more useful, we would need to be able to measure relayer subscriptions specifically, as opposed to blocks procured from relayers. This could tell us if there is a correlation between market size and connections to censoring relayers for example.  While this could be an interesting direction for future research, the role of relayers is likely to be less important with future changes to the protocol \cite{neuder2023}.

\begin{table}[htbp]
    \centering
    \normalsize
    \begin{tabular}{p{3.9cm}r}
        \toprule
        Pairwise Comparison & Strength of Correlation ($R^2$) \\
        \midrule
        Market Share vs. Clients & 0.16 \\
        Market Share vs. Relayers & 0.37 \\
        Relays vs. Clients & 0.16 \\
        \bottomrule
    \end{tabular}
\end{table}

\subsubsection{Correlation of variability across attributes for Staking Pools}

The data in table \ref{tab:r2-value-staking-pools} represents the $R^2$ values for comparison of variability between attributes in dataset A with respect to staking pools.

\begin{table}[htbp]
    \centering
    \normalsize
    \renewcommand{\arraystretch}{1.2}
    \begin{tabular}{|p{6cm}|c|}
        \hline
        \textbf{Relationship} & \textbf{$R^2$ Value} \\
        \hline
        Market Share vs. Clients & 0.01 \\ \hline
        Market Share vs. Relays & 0.02 \\ \hline
        Market Share vs. Operators & 0.0 \\ \hline
        Relays vs. Clients & 0.36 \\ \hline
        Relays vs. Node Operators & 0.11 \\ \hline
        Clients vs. Node Operators & 0.07 \\ \hline
    \end{tabular}
    \vspace{10pt}
    \caption{$R^2$ Values for Staking Pools}
    \label{tab:r2-value-staking-pools}
\end{table}

As can be seen from table \ref{tab:r2-value-staking-pools}, the only value of any significance is the $R^2$ value for Relays vs Clients.  This indicates a strong correlation between the variability in consensus clients that are used by a staking pool, and the variability in the relays that they are using.  This would suggest that staking pools that employ node operators with a higher level client of diversity also connect to multiple relayers.  While it is not surprising that staking pools with a larger market share may have multiple node operators, the results do not show a high correlation between market share and clients or relays, \textbf{indicating that the size of the staking pool does not necessarily correlate with their policies around client diversity or relayer diversity.}

\subsection{Calculating Standard and Modified Herfindahl–Hirschman Indices}

\subsubsection{Herfindahl–Hirschman Indices for Node Operators}

While the modified HHI is a useful metric for understanding general levels of correlation across node operators, we currently lack any sort of benchmark for the purposes of comparison, which poses a challenge to interpreting the results and deriving meaningful conclusions.  That being said, establishing an initial benchmark index value and measuring changes to it over time will prove useful in the long term.

In observing the \textbf{standard HHI yields a value of 66}, indicating a highly unconcentrated market of node operators.  This would suggest a very healthy level of diffusion within the node operator set, and is representative of a dataset of 2,496 records with only 22 operators that have a market share of above 1\%, and only 4 that have a market share between 2\% and 4\%.

The modified HHI yields a value of 611, which at first seems to suggest considerably higher levels of concentration., though it is important to note however, that a direct comparison between the standard HHI and modified HHI is not a practical approach to analysis.

We can expect the modified HHI to yield a much higher value than the standard HHI, which is a value that more accurately represents the fact that most node operators are using the same clients and relayers.  Given that there are only 6 consensus clients and 5 relayers with any significant market share, it is not surprising that we see this result in a population of 2,496 node operators. \textbf{The initial index value of 611 will therefore serve as a benchmark for future comparison.}, through repeated measurements over regular intervals.

\subsubsection{Herfindahl–Hirschman Indices for Staking Pools}

The standard HHI for staking pools yields a value of \textbf{1,135}, indicating a low-to-moderate level of market concentration in terms of the standard interpretation of the HHI \cite{usdoj2015}.  The modified HHI yields a value of \textbf{1,094}.  Our analysis identified 13 unique staking pools, using 6 clients and 6 relayers in numerous combinations.

As discussed with the results for node operators, a direct comparison between the standard HHI and modified HHI can be misleading.  While the standard HHI has clearly defined ranges which allow for interpretation of the results \cite{usdoj2015}, the modified HHI should be based against some baseline that is established over time, by recording and comparing results at regular intervals, (e.g. every 30 days for 12 months), in order to establish an initial benchmark.

\subsubsection{Herfindahl–Hirschman Indices for Individual Nodes}
\label{sec:herfindahl–hirschman-indices-for-individual-nodes}

As part of our analysis, we also calculated the modified HHI for individual nodes in the network in dataset B.  \textbf{The resulting modified HHI is 1,137}. Though this is a very similar result to the modified HHI for staking pools, this may be entirely coincidental.  This value can be used as benchmark against further analysis, such as the analysis conducted in section \ref{sec:ranking-the-level-of-correlation-between-attributes-in-dataset-b}, in which we rank the level of correlation between attributes in Dataset B.

We describe our approach to calculating this value in the rest of this section.  We compared nodes across 3 attributes, which were client, country and ISP.  Market share for each node was calculated by estimating the number of validators attached to the node.  For all nodes with a number of advertised attestation subnets below 64, we set the number of validators to equal to the number of attestation subnets.  We excluded any nodes that do not advertise any subnets.  For the nodes that advertise 64 or more attestation subnets, we assigned a random number of validators, based on the distribution of validators observed in dataset C.

\begin{figure}[htbp]
    \centering
    \includegraphics[width=0.95\linewidth]{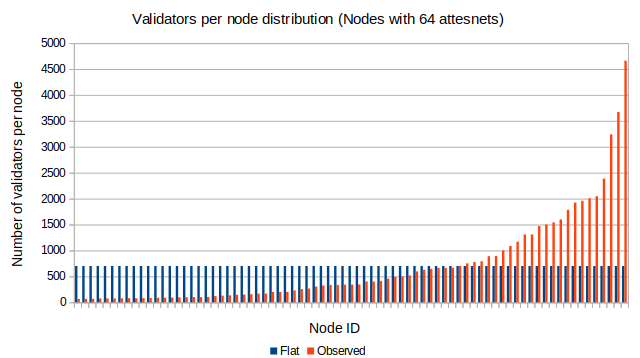}
    \caption{Validator Count Distribution Dataset C}
    \label{fig:datac}
\end{figure}

Figure~\ref{fig:datac} shows in red the observed distribution of the number of validators in the nodes with 64 attesnets for which we know their validator count (see Section~\ref{sec:data}). The blue bars represent a flat distribution if one wanted to distribute the same number of validators across all nodes equally. From here, we take the red distribution observed in these nodes, and extrapolate it to the approximately 1,000 nodes with 64 attesnet subscriptions of Dataset B, by assigning a random number of validators to each one of those nodes in such a way that it emulates the same distribution. We choose a random assignment to avoid bias towards one parameter or another. Also, we generate multiple random distributions for dataset B to further reduce the chances of adding biassed artefacts. With this, our enhanced Dataset B has the same number of validators (i.e., ~666K validators) that the network had at the time of Dataset B snapshot.

\subsection{Analysis of Individual Nodes}

\subsubsection{Measuring the correlation between attributes in Dataset B}

We present the results of analysing the correlation between individual nodes on the network in dataset B. The results are shown in table \ref{tab:pairwise-comparisons-dataset-b}, for the Chi-squared, P-value and Cramér's V value for each pairwise comparison of attributes.

As is expected, there is a very strong correlation between the country of operation and the ISP used.  This is exactly what we would expect to see given that most ISPs only serve their domestic market.

\vspace{10pt}

\begin{table}[ht]
    \centering
    \renewcommand{\arraystretch}{1.5}
    \begin{tabular}{|p{2.5cm}|c|c|c|}
        \hline
        \textbf{Comparison} & \textbf{Chi-squared} & \textbf{P-value} & \textbf{Cramér's V} \\
        \hline
        Country v ISP & 327297.96 & 0 & 0.85 \\ \hline
        Country v Client & 2426.12 & \textless 0.01 & 0.25 \\ \hline
        Client v ISP & 5883.76 & \textless 0.01 & 0.19 \\ \hline
        Subnets v Country & 5206.07 & \textless 0.01 & 0.05 \\ \hline
        Subnets v ISP & 57642.02 & \textless 0.01 & 0.15 \\ \hline
        Subnets v Client & 2176.21 & \textless 0.01 & 0.23 \\ \hline
    \end{tabular}
    \vspace{10pt}
    \caption{Pairwise Comparisons with Chi-squared Test Statistics}
    \label{tab:pairwise-comparisons-dataset-b}
\end{table}

There is a slight correlation between the number of advertised subnets and the country of operation, however, the effect size, as measured by Cramér's V of 0.048, suggests a weak association between them, which suggests that the number of validators attached to a node, and therefore the size of the node operator, is not necessarily a factor in the geographical location of the node itself.

\begin{figure}[htbp]
    \centering
    \includegraphics[width=0.95\linewidth]{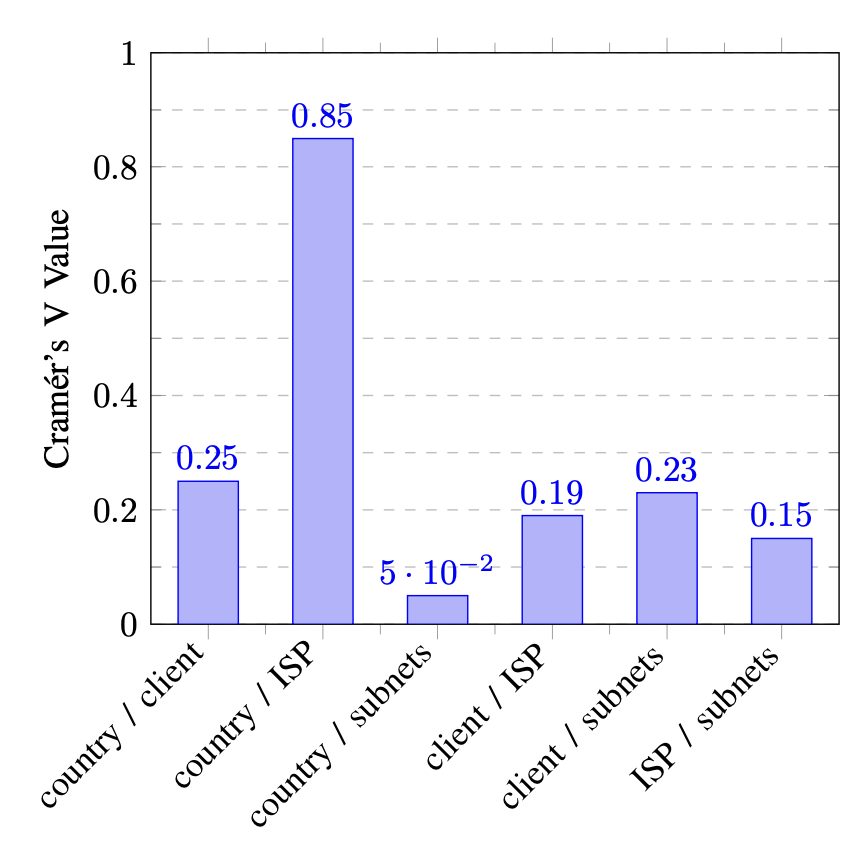}
    \caption{Effect size of correlations between attributes}
    \label{fig:relative-level-of-correlation-between-attributes}
\end{figure}

An interesting observation from the results shown in figure \ref{fig:relative-level-of-correlation-between-attributes} is that there is a relatively low correlation (or more specifically a low effect size) between the country that nodes are based in, and the number of attestation subnets that they advertise.  It is widely acknowledged that there is a strong concentration of nodes in the US and EU geographical regions, with the Rated dashboard showing 38.29\% of nodes in the US \cite{ratednetwork2024}.  The low correlation between the attributes of ``country'' and ``attestation subnets'' would indicate that nodes that have multiple validators attached to them, and therefore likely to be operated by node operators serving staking pools, may be more geographically distributed.  Available data from Lido suggests that they have a slightly wider geographical distribution of nodes \cite{lido2024}.

The number of advertised attestation subnets and the consensus clients used seem to be significantly associated ($\chi$²=2176.21, p\textless 0.01) with a moderate effect size (V = 0.233), which suggests some sort of correlation. Similarly, there is a significant association between ISP and consensus client ($\chi$²=5883.76, p\textless 0.001) with a moderate effect size (V=0.186). \textbf{These indicators of correlation may potentially indicate preferences among larger node operators, who may have a bias toward high performance and low latency operational configurations.}

\subsection{Correlation between node operator market share and clients and relayers}

\subsubsection{Operator market share and clients used}

\begin{figure}[htbp]
    \centering
    \includegraphics[width=0.95\linewidth]{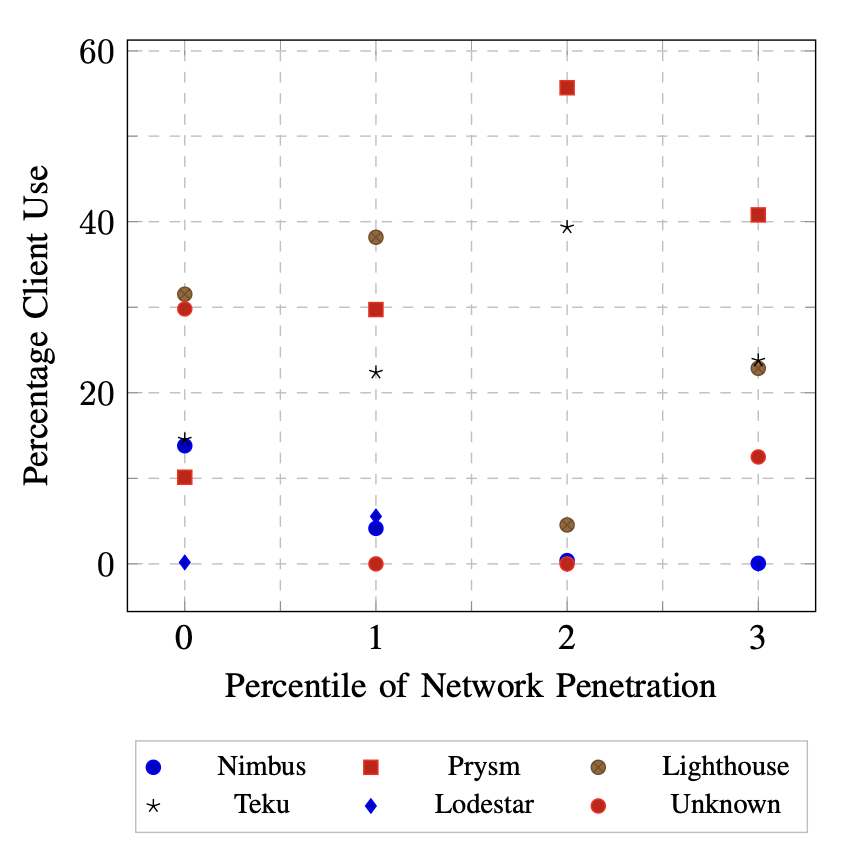}
    \caption{Average Client Percentage by Node Operator Market Share}
    \label{fig:average_client_percentage_by_percentile}
\end{figure}

The scatter plot in figure \ref{fig:average_client_percentage_by_percentile} shows the correlation between the market share of node operators in Dataset A, and the average percentage of each consensus client that node operators in the respective market share percentile use.

The 6 main consensus clients are listed in the legend below the chart.  The results displayed in the chart roughly align with the market share of the various clients, however, the average percentage of client use changes for different clients as the operator's market share increases.  \textbf{As can be observed, the Prysm client has a larger average percentage of use among larger node operators whereas Nimbus and Lighthouse sees less average percentage use as the size of the node operator increases.}

This observed trend could potentially be attributed to the fact that Lighthouse has been reported to be highly stable and performant \cite{ranjan2023}, which would make it a preferred choice for larger node operators, whereas Nimbus requires less hardware resources, possibly making it more appealing to solo stakers.

The data broadly agrees with the overall consensus client distribution, which has Prysm and Lighthouse clients at approximately 30\% and 40\%, with other clients having lower shares.


\subsubsection{Operator market share and relayers used}

\begin{figure}[htbp]
    \centering
    \includegraphics[width=0.95\linewidth]{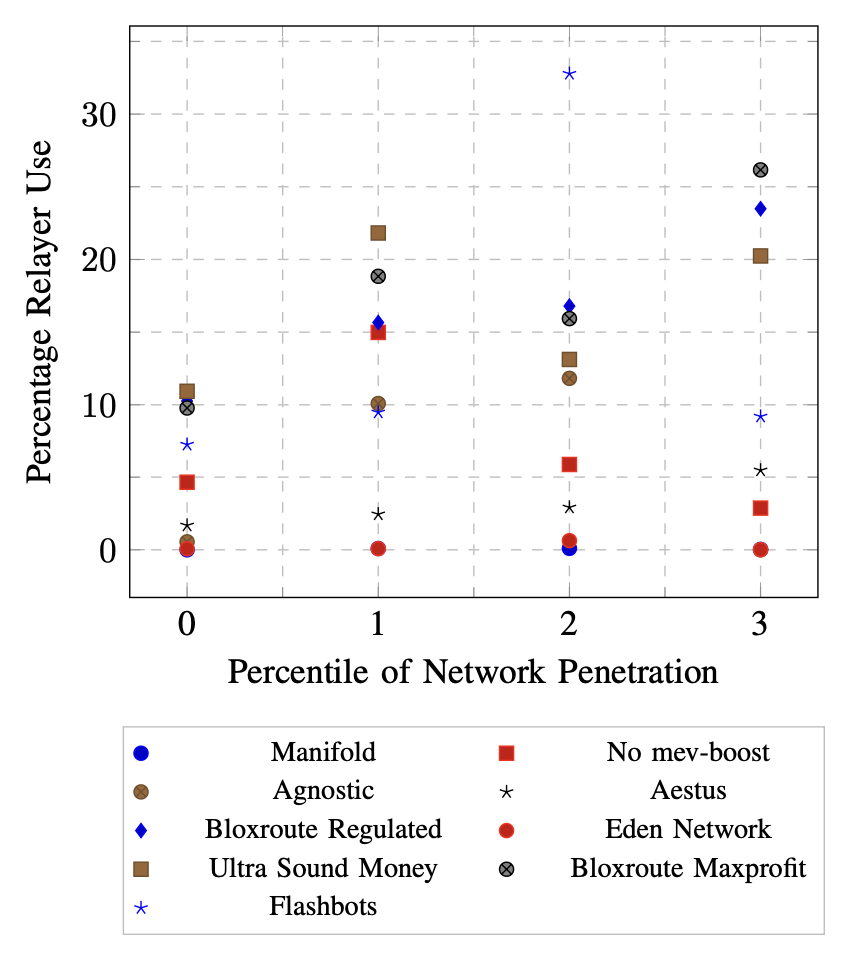}
    \caption{Average Relayer Percentage by Node Operator Market Share}
    \label{fig:average_relayer_percentage_by_percentile}
\end{figure}

The scatter plot in figure \ref{fig:average_relayer_percentage_by_percentile} shows the correlation between the market share of node operators and the average percentage of each relayer that node operators in the respective market share percentile use.  As can be observed from the chart, there is an increase in variability as the market share / size of the node operator increases.  This is in line with previous analysis that showed an $R^2$ value of 0.37 for Market Share vs. Relayers. Again, this potentially indicates that larger node operators may connect to more relayers, but this requires further analysis.

\subsection{Ranking the level of correlation between attributes in Dataset B}
\label{sec:ranking-the-level-of-correlation-between-attributes-in-dataset-b}

\begin{figure}[htbp]
    \centering
    \LARGE \textcolor{bluebullet}\textbullet\ \normalsize Subnets 
    \LARGE \textcolor{redbullet}\textbullet\ \normalsize Country 
    \LARGE \textcolor{yellowbullet}\textbullet\ \normalsize ISP 
    \LARGE \textcolor{greenbullet}\textbullet\ \normalsize Client
    \includegraphics[width=1\linewidth]{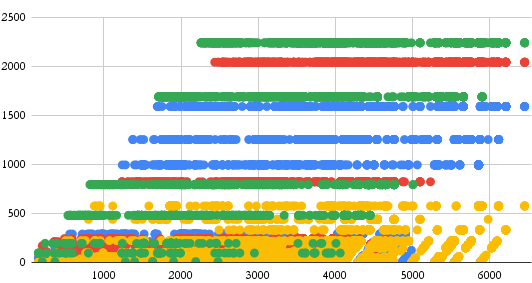}
    \caption{Ranking the correlation between attributes in Dataset B}
    \label{fig:attribute-correlation-ranking-dataset-b}
\end{figure}

Our analysis attempts to rank the level of correlation between the various attributes in dataset B, pertaining to individual nodes on the network.  The results are visualized in figure \ref{fig:attribute-correlation-ranking-dataset-b}, as a scatter-plot chart.

These charts visualize how often the values for client, country, ISP, and subnets for each node occurs in the dataset.

Each node is displayed along the horizontal axis, and is ranked by sum of the occurrences within the dataset for the value for each attribute of that node, with values to the right showing highers sums.  Each node is placed along the horizontal axis and has four vertically aligned markers which correspond to the four attributes of country, client, ISP, and subnets.  Each marker represents the number of times the value for that attribute occurred in the dataset.

We can see that the most common values that exist in the dataset tend to be in the client, country, and subnet variables.  When we examine the nodes with highest aggregate amount of common values across all attributes, we observe that these nodes share values most often in the client, country and subnet fields.

While these trends do not point to any linear relationship (where the more subnets a node advertises, the more likely they are to be in a certain country for example), it does suggest that there is some concentration within the distribution of values for these attributes, and that there is potentially some correlation.


We observe several horizontal bands within the data in figure \ref{fig:attribute-correlation-ranking-dataset-b}.  These bands represent specific attribute values that are common within the dataset.  These horizontal bands are more dense around 20,000 to 50,000 on the horizontal axis, and appear in near isolation toward the right hand side of the scatter-plot, above 50K (i.e. with fewer other values for those nodes).  This points to a cluster of nodes that share common values across a number of attributes.  The specific values for this cluster listed as follows:

The three distinct blue horizontal bands in the middle of the subgraph represent the nodes that advertise either 1, 2, or 64 attestation subnets, which collectively accounts for approximately 70\% of the dataset. This shows two distinct groups of solo stakers and larger node operators within the data.

The two green horizontal bands near the top of the chart represent the consensus client market shares of Lighthouse and Prysm, which both have 40\% and 30\% shares of the market respectively.

The distinct red horizontal band near the top of the chart represents nodes that are based in the US, which accounts for 37.1\% of the dataset, followed by another distinct red horizontal band below the center of the vertical range, which represents nodes based in Germany, which accounts for 14.9\% of the dataset.  This corresponds broadly with the observed geographical centralization in network nodes in general for both consensus and execution nodes \cite{nodewatch2024}.

The top right hand quadrant of the chart identifies a cluster of nodes, which have the highest aggregate amount of common values for each attribute.  \textbf{These are nodes that are based in the US, are running either Lighthouse or Prysm, and are advertising either 1 or 2 attestation subnets, suggesting they are operating as solo stakers}.  While there is a definite cluster of such nodes in the data, they are constrained to the far right hand side of the graph, suggesting a somewhat small effect size, which aligns with the results from section VI, which concluded a low effect size for the correlation between country and subnets.  This suggests that this is not currently a concerning trend, but is potentially a trend worth monitoring.

\subsection{Ranking the level of correlation between attributes in Dataset C}

\begin{figure}[hbp]
    \centering
    \LARGE \textcolor{bluebullet}\textbullet\ \normalsize Validators 
    \LARGE \textcolor{redbullet}\textbullet\ \normalsize Client 
    \LARGE \textcolor{yellowbullet}\textbullet\ \normalsize Country 
    \LARGE \textcolor{greenbullet}\textbullet\ \normalsize ISP
    \includegraphics[width=1\linewidth]{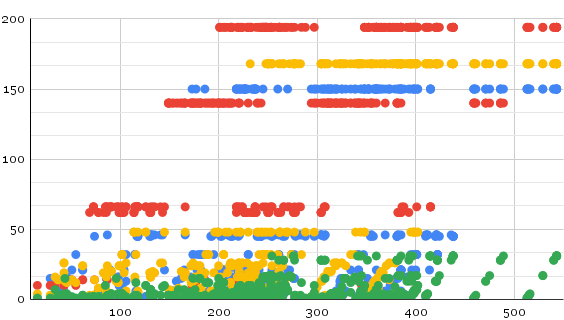}
    \caption{Ranking the correlation between attributes in Dataset C}
    \label{fig:attribute-correlation-ranking-dataset-c}
\end{figure}

The same analysis was applied to dataset C, the results of which are displayed in figure \ref{fig:attribute-correlation-ranking-dataset-c}.  The scatter-plot displays nodes that have more overall correlation between attributes to the right.  As can be observed, the highest level of correlation is found in the top right quadrant of the chart, where there are clear levels of correlation between client, country and the number of validators attached to the node. 

\vspace{4pt}

It is worth noting that while including the validator count as a variable in this analysis seems counter-intuitive due to the fact that is a continuous variable, as opposed to categorical like the other attributes, there are broad categories within the data, such as solo stakers for example, and many medium sized stakers, all of whom share similar numbers of validators, making the variable suitable for consideration as a categorical variable.

The correlation between validators and country is based on the number of validators nodes have in common, which does not necessarily indicate a linear relationship, whereby an increase in the number of validators would increase the probability of being in a certain country.  The distribution of the number of validators per node has a mode of 1 and median of 4, \textbf{forming a bottom heavy distribution that would suggest that geographical concentration of nodes may be driven more by solo stakers, though only to a relatively low degree}.

This may explain the ostensibly contradictory results in table \ref{tab:pairwise-comparisons-dataset-b}, which shows a relatively low level of correlation between advertised subnets and the country of operation.

\section{Conclusion}
\label{sec:conclusion}

Overall the results identified a number of broad observable trends:

Larger node operators tend to propose blocks from more relayers then smaller operators and solo stakers, who tend to propose blocks from only one or a small set of relayers.  This can potentially be explained by solo stakers winning less blocks, but also might be an indication that solo stakers connect to fewer relayers, though without more data it is difficult to determine.

There is a moderate level of correlation between the ISP used and the number of attestation subnets advertised, which may suggest that nodes that have multiple validators attached, usually larger node operators, are often using the same public cloud providers, such AWS or Azure.

The Prysm consensus client has a larger average percentage of use among larger node operators, whereas Nimbus and Lighthouse have less. This observed trend could potentially be attributed to the fact that Lighthouse has been reported to be highly stable and performant, making it a preferred choice in for larger node operators, whereas Nimbus requires less hardware resources, possibly making it more appealing to solo stakers \cite{ranjan2023}.

We observe indications that there is a concentration of geographical location of validator nodes, which corresponds broadly with the observed geographical centralization of all nodes on the network, i.e. validators nodes and non-validator nodes together. However, the discrepancy between the high correlation of geographical location and number of attached validators (see figure \ref{fig:attribute-correlation-ranking-dataset-c}), with the low level of correlation between geographical location and attestation subnets advertised, (see \ref{fig:relative-level-of-correlation-between-attributes}), suggests that geographical concentration may in fact be driven more by smaller node operators and solo stakers, rather than by larger node operators.  This could be due in part to geographical diversity policies of staking pools such as Lido \cite{vanom2024}.

In summary, it would appear that there is both a slightly lower degree of client diversity, as well as a higher concentration of nodes in public cloud data centres, that can be attributed to larger node operators.  However, larger node operators may tend to increase the diversity in mev-boost relays used, and there are indications that they may favor geographical diversity of nodes.

\section{Discussion}
\label{sec:discussion}

Observing the percentage breakdown of certain aspects of the node distribution within the network can reveal important insights.  The insights include any unequal distribution of client software used by nodes, or unequal distribution of geographical region of operation.  Both concerns may have implications for the health of the network.  However, by looking at the level of correlation between staking pools and node operators on the network, we can reveal useful clues as to where the drivers of any concentration occur.  For example, observed trends from our results suggest that larger node operators are potentially driving a concentration in certain areas (with the exception of geographical distribution as described in section \ref{sec:conclusion}).  While the degree to which this concentration occurs is not currently concerning, it would be prudent to measure changes to these levels of concentration over time.

Further to identifying potential drivers of concentration through looking at patterns of correlation, there is also a benefit in developing a standardized index for the measurement of decentralization.  While it can be useful to look at individual aspects of a network, such as geographical distribution of nodes, distribution of nodes by public cloud provider, distribution of client software used etc., it can be a challenge to agree on the overall effective level of decentralization when these various measurements are combined.  This can be all the more challenging when attempting to measure any changes in the overall effective level of decentralization over time.  Similar challenges in traditional economic sectors led to the development of economic indices such as the Gini Index and Herfindahl–Hirschman Index.  For the same reason, it is useful to have a standard index that can be applied to cryptocurrency networks to derive a high-level, comparative measurement.

The modified HHI proposed in this paper can be applied to other areas as well.  In the future the same model can be applied to Layer 2 rollups as they start to decentralize their sequencers.  When this starts to happen, we may have a scenario where some L2s will have a large market share, but will be fully centralized, where other L2s will have a smaller market share and will have fully decentralized sequencers and/or provers.  In this scenario, simply measuring decentralization via observing the relative market share of each L2 is insufficient to capture the effective level of decentralization within the ecosystem.  There may emerge several nuanced factors that should be considered within a decentralized sequencer set, or a shared sequencer network, such as governance parameters, geographic distribution of nodes, or the jurisdiction of headquarters of companies of validator node operators.  It may be the case that some L2 have a rich client diversity within their validator set whereas others do not.  These considerations can be used in conjunction with existing models as well as existing methodologies and frameworks \cite{l2beat2024}.

It is worth noting that it is currently quite challenging to collect accurate data on the correlation between validators and nodes, and node operators.  This is in part by design, as the protocol is designed in such a way as to protect the identity, including the IP address, of validator nodes.  This provides a level of protection from DDOS attacks, which could have economic consequences if the validator is known to be a proposer and is not able to deliver a block to the network.  However, this also makes it challenging to measure the level of correlation between nodes, which can point to the drivers of concentration in certain areas. Current data leverages a mixture of self-reporting and various sources of proxy data, which is cumbersome and sub-optimal. An interesting future direction of research would be to identify mechanisms that can potentially be employed to implement some form of protocol-level inspection that would allow some certainty over metrics used to calculate the level of decentralization in the network.

\vspace{8pt}

\section*{Acknowledgments}

The authors of the paper would like to express their gratitude to the Lido Ecosystem Grants Organization for their support of this work. Also, we would like to thank the Ethereum Foundation for their time and insights with respect to developing an index using a correlation factor.

We would like to thank Elias Simos at rated.network and Isidoros Passadis at Lido for generously dedicating time to reviewing the paper and providing advice and feedback. The authors would also like to thank Alvaro Revuelta and the MigaLabs team for their support in obtaining this data, and Rob Dawson at Consensys for his continual encouragement and support.

\vspace{12pt}

\printbibliography

@article{wu2020coefficient,
  title={A Coefficient of Variation Method to Measure the Extents of Decentralization for Bitcoin and Ethereum Networks.},
  author={Wu, Keke and Peng, Bo and Xie, Hua and Zhan, Shaobin},
  journal={Int. J. Netw. Secur.},
  volume={22},
  number={2},
  pages={191--200},
  year={2020}
}

@inproceedings{gupta2018gini,
  title={Gini coefficient based wealth distribution in the bitcoin network: A case study},
  author={Gupta, Manas and Gupta, Parth},
  booktitle={Computing, Analytics and Networks: First International Conference, ICAN 2017, Chandigarh, India, October 27-28, 2017, Revised Selected Papers 1},
  pages={192--202},
  year={2018},
  organization={Springer}
}

@article{gochhayat2020measuring,
  title={Measuring decentrality in blockchain based systems},
  author={Gochhayat, Sarada Prasad and Shetty, Sachin and Mukkamala, Ravi and Foytik, Peter and Kamhoua, Georges A and Njilla, Laurent},
  journal={IEEE Access},
  volume={8},
  pages={178372--178390},
  year={2020},
  publisher={IEEE}
}

@article{lee2021dq,
  title={DQ: Two approaches to measure the degree of decentralization of blockchain},
  author={Lee, Jaeseung and Lee, Byungheon and Jung, Jaeyoung and Shim, Hojun and Kim, Hwangnam},
  journal={ICT Express},
  volume={7},
  number={3},
  pages={278--282},
  year={2021},
  publisher={Elsevier}
}

@article{brown2023measuring,
  title={Measuring the Concentration of Control in Contemporary Ethereum},
  author={Brown, Simon},
  journal={arXiv preprint arXiv:2312.14562},
  year={2023}
}

@misc{beaconchain2024,
  title={Open Source Ethereum Blockchain Explorer - Beaconcha.in - 2024},
  author={beaconcha.in},
  url={https://beaconcha.in/},
}

@misc{nodewatch2024,
  author={ChainSafe Systems Inc.},
  title={NodeWatch - ETH2 Node Analytics},
  url={https://nodewatch.io/},
}

@misc{dune2024,
  author={Hildobby},
  title={ETH Stakers},
  url={https://dune.com/queries/2394100/3928083},
}

@misc{wahrstatter2023,
  author={ETHGlobal},
  month={12},
  title={{Toni Wahrstätter <unk> Exploring censorship across the PBS stack}},
  year={2023},
  url={https://www.youtube.com/watch?v=WcJlseuhbX8},
}

@misc{clientdiversity2024,
  title={Client Diversity | Ethereum},
  url={https://clientdiversity.org/#distribution},
}

@misc{vanom2024,
  title={Lido on Ethereum Validator and Node metrics},
  url={https://app.hex.tech/8dedcd99-17f4-49d8-944e-4857a355b90a/app/3f7d6967-3ef6-4e69-8f7b-d02d903f045b/latest},
}

@misc{OECD2021,
  author={Bascunana-Ambros, Patricia and Capobianco, Antonio and Zigelski, Sabine and Meester, Wouter},
  title={Methodologies to Measure Market Competition},
  institution={OECD Competition Committee},
  type={Issues Paper},
  year={2021}
}

@article{ranjan2023,
  author={Ranjan, Pooja},
  month={3},
  title={Exploring Ethereum’s client Ecosystem - Ethereum Cat Herders - Medium},
  year={2023},
  url={https://medium.com/ethereum-cat-herders/exploring-ethereums-client-ecosystem-afc9affa84dd},
}

@misc{l2beat2024,
  title={Methodology and Framework},
  url={https://gov.l2beat.com/c/methodology-and-framework/5},
}

@article{pearson1895,
  author={Pearson, Karl},
  title="{Note on Regression and Inheritance in the Case of Two Parents}",
  journal={Proceedings of the Royal Society of London Series I},
  year=1895,
  month=jan,
  volume={58},
  pages={240-242},
  adsurl={https://ui.adsabs.harvard.edu/abs/1895RSPS...58..240P},
}

@misc{usdoj2015,
  month={6},
  title={{Horizontal Merger Guidelines (08/19/2010)}},
  author={The United States Department of Justice},
  year={2015},
  url={https://www.justice.gov/atr/horizontal-merger-guidelines-08192010#5c},
}

@article{akoglu2018user,
  title={User's guide to correlation coefficients},
  author={Akoglu, Haldun},
  journal={Turkish journal of emergency medicine},
  volume={18},
  number={3},
  pages={91--93},
  year={2018},
  publisher={Elsevier}
}

@misc{ratednetwork2024,
  title={{Network Overview | Ethereum Ethereum Mainnet}},
  url={https://www.rated.network/overview?network=mainnet&timeWindow=1d&rewardsMetric=average&geoDistType=all&hostDistType=all&soloProDist=stake},
}

@misc{lido2024,
  author={Lido},
  month={2},
  title={{VaNOM: Lido on Ethereum Validator \& Node metrics}},
  year={2024},
  url={https://app.hex.tech/8dedcd99-17f4-49d8-944e-4857a355b90a/app/3f7d6967-3ef6-4e69-8f7b-d02d903f045b/latest?selectedStaticCellId=28257761-e81d-4b2b-8fa8-a0c3d5dd07de},
}

@misc{neuder2023,
  author={Neuder, Mike},
  month={8},
  title={{Relays in a post-ePBS world}},
  year={2023},
  url={https://ethresear.ch/t/relays-in-a-post-epbs-world/16278},
}

\end{document}